\documentclass[a4paper, 10pt, onecolumn, conference]{ieeeconf}      
\usepackage{FG2023}

\FGfinalcopy 

\usepackage{graphicx}
\usepackage{tikz}
\usepackage{comment}
\usepackage{float}
\usepackage{amsmath,amssymb} 
\usepackage{hyperref}
\usepackage{color}
\usepackage{booktabs}
\usepackage{arydshln}

\def\FGPaperID{100} 

\title{\LARGE \bf
SignNet: Single Channel Sign Generation using Metric Embedded Learning
}

\author{\parbox{16cm}{\centering
		{\large Tejaswini Ananthanarayana$^1$, Lipisha Chaudhary$^2$ and Ifeoma Nwogu$^2$}\\
		{\normalsize
			$^1$ SONY Corporation, USA\\
			$^2$ Department of Computer Science and Engineering,  University at Buffalo, NY, USA}\\
		\texttt{ta2184@rit.edu;\{lipishan,inwogu\}@buffalo.edu}} 
}

\begin{document}
	
%
\maketitle

\begin{abstract}
	A true interpreting agent not only understands sign language and translates to text, but also understands text and translates to signs. Much of the AI work in sign language translation to date has focused mainly on translating from signs to text. Towards the latter goal, we propose a text-to-sign translation model, SignNet, which exploits the notion of similarity (and dissimilarity) of visual signs in translating. This module presented is only one part of a dual-learning two task process involving text-to-sign (T2S) as well as sign-to-text (S2T).  We currently implement SignNet as a single channel architecture so that the output of the T2S task can be fed into S2T in a continuous dual learning framework. By single channel, we refer to a single modality, the body pose joints. 
	
	In this work, we present SignNet, a T2S task using a novel metric embedding learning process, to preserve the distances between sign embeddings relative to their dissimilarity. We also describe how to choose positive and negative examples of signs for similarity testing. From our analysis, we observe that metric embedding learning-based model perform significantly better than the other models with traditional losses, when evaluated using BLEU scores. In the task of gloss to pose, SignNet performed as well as its state-of-the-art (SoTA) counterparts and outperformed them in the task of text to pose, by showing noteworthy enhancements in BLEU 1 - BLEU 4 scores (BLEU 1: 31 $\rightarrow$ 39; $\approx$26\% improvement and BLEU 4: 10.43 $\rightarrow$ 11.84; $\approx$14\% improvement) when tested on the popular RWTH PHOENIX-Weather-2014T benchmark  dataset
	
\end{abstract}

\section{Introduction}
\label{sec:intro}
Two-way voice-controlled systems such as Alexa by Amazon, Siri by Apple, Bixby by Samsung, etc.~are becoming more popular with recent advances in technology. These systems have proven extremely beneficial for hearing and speaking individuals, but not necessarily so for the Deaf-and-Hard-of-Hearing (DHH) community. For this and many other similar reasons, sign language analysis is becoming a more prevalent research area in the AI  community.

But much of the AI work to date has focused primarily on translating sign language to text (which can be readily extended to speech). Unfortunately, this again puts the advantage on the side of the hearing-centric rather than the DHH, where they receive information in their own natural language. In this work, we present a model that when coupled with some of the current state-of-the-art sign-to-text (then to voice) technologies, can facilitate two-way sign language analysis. The work presented here is only one half of an overall strategy, 
involving the dual-learning of two complementary tasks -  text-to-sign (T2S) and sign-to-text (S2T).  We currently implement SignNet as a single channel architecture so that the output of the T2S task can be seamlessly fed into S2T in a continuous dual learning framework. By single channel, we refer to a single modality, the body pose keypoints. Although several different multichannel frameworks (involving pose, optical flow, etc.) have been proposed, they cannot as readily feed into a dual-learning scheme, hence our focus at this time on single channel generation.

\begin{figure}[t]
	\centering
	\includegraphics[width=1.0\linewidth]{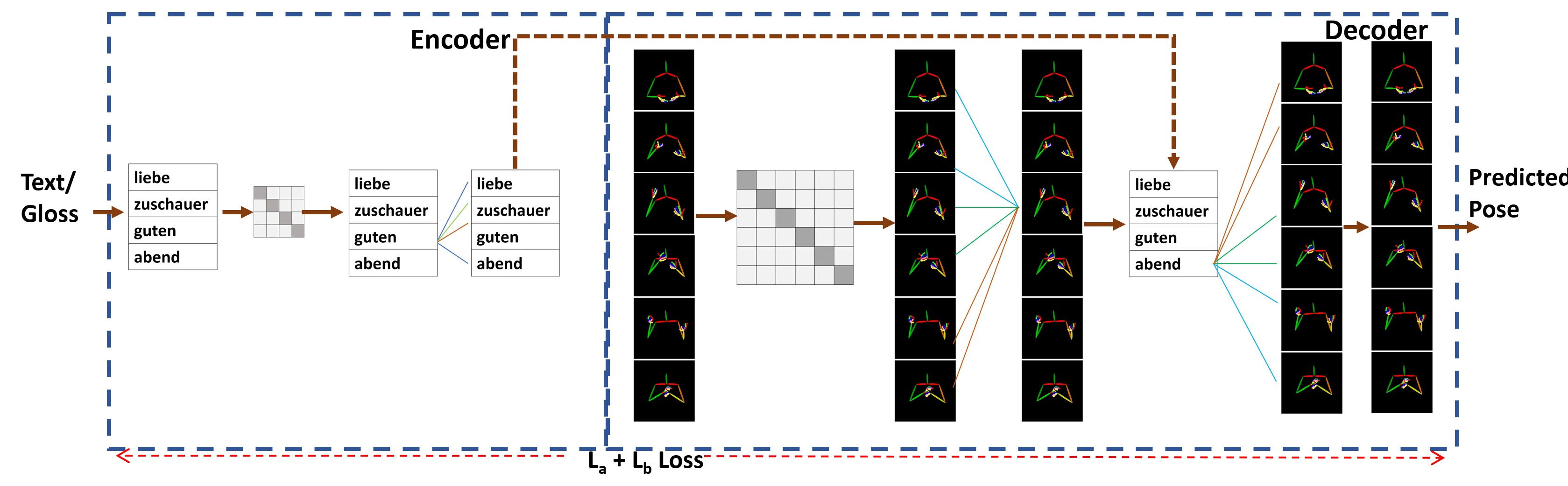} 
	\caption{An overview of our proposed SignNet architecture, showing the text encoder learn a sign embedding; a sign similarity learning mechanism is then implemented in the embedding space, resulting in more precise generation of sequences of sign poses via the decoder. $L_{a}, and L_{b}$ are the losses propagated backwards through the network. Details of the architecture are described in Section \ref{method:signn}.}\label{fig:signnet}
\end{figure}

\textit{The main intellectual merit of this paper is the introduction of a learning mechanism, consisting of different aggregated losses, including a novel similarity metric-based loss, useful for efficient single channel pose generation during text to sign language interpretation.
}

The main societal benefit of this work is the burgeoning of a quintessential interpreting agent that not only understands sign language and translates to text, but also understands  text and translates to signs. Our work plays a role in the function of latter half of such an agent. Such an end-to-end sign language interpretation agent will prove useful, not only to the hearing-centric population (as is the case for most SoTA AI models for continuous sign language translation) but also to the DHH community. This is important as it allows both the hearing and the DHH  to converse freely with each other, in their own preferred natural language. 
The model we present takes as input text phrases, and translates these to signs in form of poses (more on pose as a representation for signs is discussed in Section \ref{method:signn}); we evaluate the generated poses by verifying the translations using a  sign-to-text model.

\subsubsection{Importance of context in sign language translation}\label{sec:context}
Sign language has some unique linguistic aspects that prove very challenging for automated sign translation, especially those that involve the use of the 3D space around the signer (referred to as the ``signing space'') \cite{huenerfauth2012effect}. While conversing, a signer will often associate places, people, and different entities with specific 3D locations close to her body. For example, a signer could finger-spell the name \emph{J.a.n.e.t.}, the first time that person is mentioned in the conversation. The signer would now point to a location around her body and a spatial reference has now been created associating the person named \emph{Janet} to that 3D location. In the course of the conversation, for every ensuing instance in which the signer wants to refer to \emph{Janet}, she will point to that location, the spatial reference point. In some instances, the signer may instead aim her gaze or tilt her head at the 3D location. As the conversation progresses and \emph{Janet} is no longer in context, that reference point diminishes. 

As another SL linguistic example, a signer may create spatial reference points, one on her left side for the subject and another on her right side for the object of the verb. In the course of the conversation, if there is a need to compare and contrast the subject and object, the signer may twist at the waist and aim her torso to the right when discussing the object and then similarly twist to the left for the subject. This seeming exaggerated movement stresses which of the entities is being discussed and this is called \emph{contrastive role shifting}.

There are significantly many more such 3D linguistic structures in sign language, where the context of the conversation is imperative in fully comprehending and translating the signs. With so many complicated constructs in sign language grammar, the current states of sign language recognition, translation and generation are still in the very infancy  stages of research.


\section{Related Works on Sign Generation}
Although the domain of machine-based language translation has been quite well studied, using deep learning  neural machine translation (NMT) \cite{bahdanau2016neural} models is still very active research area. Sign language translation is an extension of such language translation models, with the additional complexity of incorporating video processing to assist with the fact that sign languages are visio-temporal languages. Several different approaches such as  automated human-like avatars, conditional video generation models using generative adversarial networks (GANs), and video-based NMT approaches have been applied to address the text-to-sign language generation problem. 

Other methods have been developed to assist deaf and hard-of-hearing individuals communicate more effectively with hearing individuals employ visual aids which construct signs based on manufactured phrases \cite{Glauert2006VANESSAA}, \cite{KARPOUZIS200754}, \cite{McDonald2015AnAT}. Other approaches for generating signs include the avatar creation method \cite{10.1007/978-3-642-23974-8_13} where the authors used a popular character animation system to created a signing avatar using  already pre-identified signs. The goal here was to explore the technical feasibility of avatars for sign generation, and also develop evaluation methods. Ebling et al. \cite{ebling-huenerfauth-2015-bridging} furnished the avatars with a more humanized form, removing several ambiguities associated with employing sign language systems for effective communication.

More recently, neural network models have been used to improve the quality of sign creation. Stoll et al. created a system called Text2Sign in which Generative Adversarial Networks (GANs) were used to generate  sign videos. Zelinka et al. \cite{Zelinka_2020_WACV} created signs using a skeletal model based on the Openpose \cite{openpose_1} framework. Each of these models generated a sign for each word in a phrase, however, Saunders et al. \cite{progressivetransformer} improved on this by generating 3D continuous signs using gloss for human posture generation. The same group of authors created a 3D multi-channel sign language generation approach using transformers and mixture density networks \cite{progressive}. Other multi-channel works include\cite{SaundersBMVC20}. These sign productions were developed and tested using the RWTH PHOENIX-Weather-2014T (PHOENIX14T) German sign language dataset \cite{KOLLER2015108}, which has now become a benchmark for continuous sign language analysis. An extensive review of sign language generation approaches up to 2021, was presented by Rastgoo et al., \cite{DBLP:journals/corr/abs-2103-15910}.

Beyond this, in 2022, Viegas et al. \cite{viegas2022including} proposed a sign generation model that included the use of facial expressions (the first of its kind), to capture the grammatical and  semantic  functions of sign language. And although they successfully showed that the inclusion of facial expressions improved the sign generation results, and this is an advancement in a more comprehensive sign representation, their best performing models were far below the SoTA values. 
%
%
%

\section{SignNet - Generating signs from spoken language sentences}
\label{method:signn}
Spoken language text phrases are fed to the text-to-pose network as shown in Figure \ref{fig:signnet}.  For these initial texts, we obtain their word embeddings and pass them as inputs to the encoder. Positional information is added to the word embeddings to maintain the word order information. Inter-dependencies between different words of a sentence are learned by performing multi-head attention and this information is then passed on to the decoder.  

In the decoder, 3D poses are used as the output frame features. We first obtain 2D OpenPose joints \cite{openpose_1} for each frame of the signing video and then obtain 3D pose features \cite{Zelinka_2020_WACV} from these 2D OpenPose joints. Keeping pose generation in mind we mainly target the upper body joints - hand joints, finger joints, and body joints as far down as the trunk. All these constitute 50 joint locations, thus giving us a vector of size $150$ for each frame. The input can be denoted as $POSE=\{[(x_{0_{1}}, y_{0_{1}}, z_{0_{1}}), \dots, (x_{149_{1}}, y_{149_{1}}, z_{149_{1}})], \dots, \allowbreak [(x_{0_{N}}, y_{0_{N}}, z_{0_{N}}), \dots, (x_{149_{N}}, y_{149_{N}}, z_{149_{N}})]\}$, where $(x,y,z)$ represent the 3D joints and $N$ is the number of frames. For this work, we used the already extracted 3D features from the Progressive Transformer model.

To retain the input ordering information in our model, we implement positional encoding  on input joints representation. Since sign language heavily depends on the context, we learn temporal dependencies by co-relating all the frames with respect to a single frame, continuously for all the frames.

\begin{figure*}[t]
	\centering
	\includegraphics[width=0.99\textwidth]{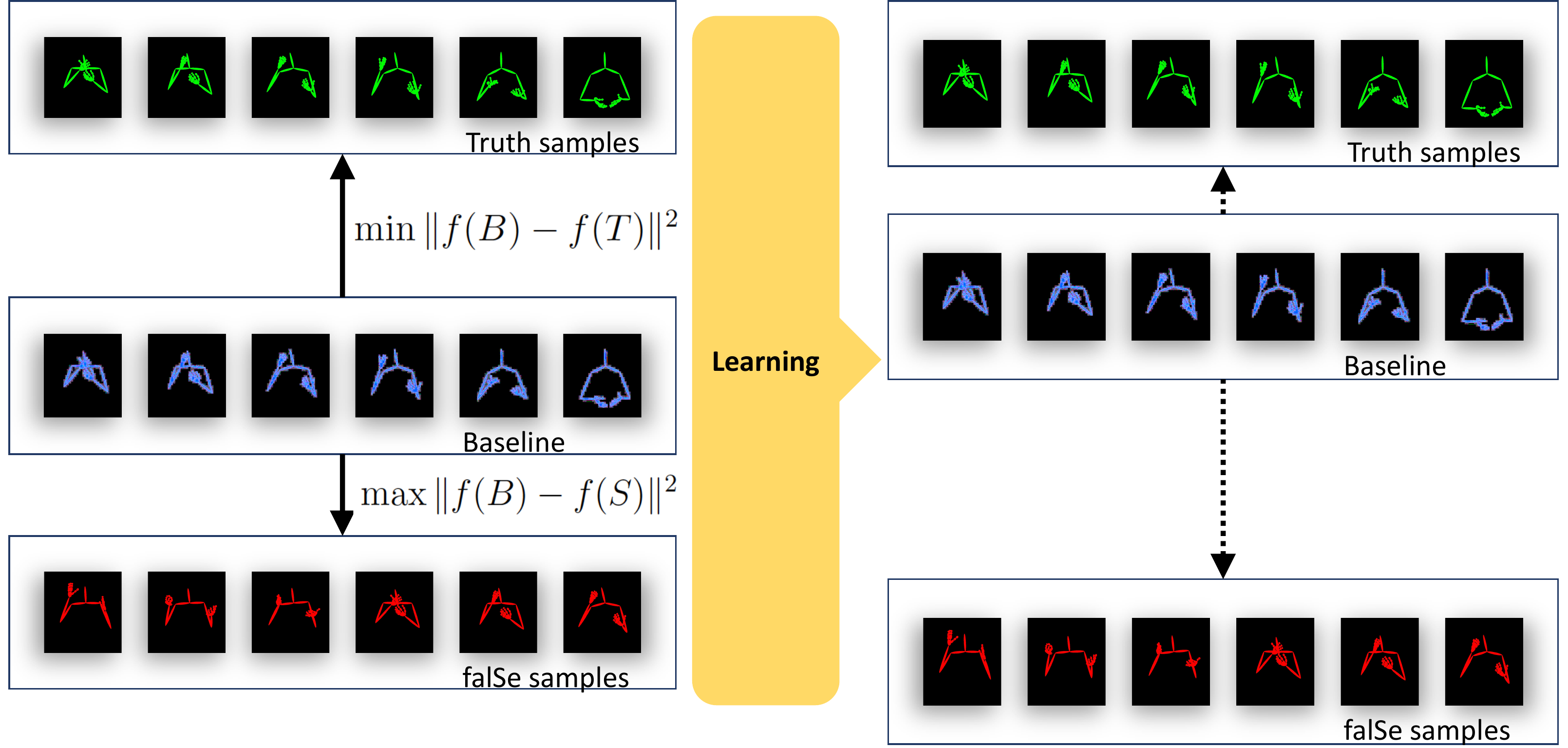} 
	\caption{The loss based on sign similarity metrics minimizes the distance between a \underline{B}aseline sign (ground-truth) and  the \underline{T}ruth-like sign (the prediction for a sample-under-investigation),  while maximizing the distance between the \underline{B}aseline and a fal\underline{S}e sample(
		a different truth sign selected from the same batch as the sample-under-investigation). The LHS shows the the sets of samples before training and the RHS shows how similar signs have become closer, while the different ones have become farther apart in the embedding space.}
	\label{fig:triplet}
\end{figure*}

We learn the context between frames by performing the scaled dot-product attention whose output - the vector V, is weighted by the queries Q and keys K. To avoid exploding values after the dot product, we scale by $\sqrt{d}$ as in \cite{transformer}. Finally, to retain the context information relevant to each frame, Softmax activation ($Softmax(\frac{QK^{T}}{\sqrt{d}})V$) is applied  on the frames V. We learn the mapping between the frames and words by taking the context information from the encoder and performing a scaled dot product with word-based attention. These learned embeddings are then passed onto a linear feed-forward network 
to predict continuous poses.

\subsection{Metric Embedded Learning for Pose Similarity}\label{sec:triplet}

We are interested in ensuring that the pose-based signs that the SignNet architecture predicts are as similar as possible to the ground-truth signs, and as distant as possible to other signs in the same training batch. Figure. \ref{fig:triplet} provides a visualization of the desired mechanism.

To accomplish this, we have:
\begin{eqnarray}
\underbrace{\|f(B) - f(T)\|^2}_{d(B,T)} -  \underbrace{\|f(B) - f(S)\|^2}_{d(B,S)} \leq 0\label{eqn:triplet1}
\end{eqnarray}

\noindent where $B$ is a baseline sign, $T$ is a truth sign required to be as similar to $B$ as possible and  $S$ is a false sign (not as similar to the baseline); $d(.)$ is the distance function.

To avoid the trivial solution where our function $f(.)$ will produce zero, or one where $f(B)=f(T)$, we introduce a margin to impose a stronger constraint, similar to \cite{FaceNet}. The resulting distance function $d(B,T,S)$ is  given as:
\begin{eqnarray}
d(B,T,S) = \max ( \hspace{1mm}(d(B,T) - d(B,S)+ \alpha), 0\hspace{1mm})\label{eqn:triplet3}
\end{eqnarray}

\noindent We refer to the loss derived based on this distance as the \emph{pose similarity metric-based loss function}, in Equation \ref{eqn:triplet4}.

\paragraph*{Choosing the similarity metric samples:} While any random choice can readily satisfy $d(B,T)+ \alpha \leq d(B,S)$, the underlying neural network will simply not learn if it gets it right too many times. But if the choice of samples is done such that $d(B,T) \approx d(B,S)$, the network is forced to work hard to learn the differences. This seemingly simple choice significantly increases the efficiency of the learning algorithm. Hence, we select our samples as follows:\smallskip

Consider a batch $\|B\| = 4$, where we are interested in calculating the similarity loss for the first sample $i=1$. The baseline here is the ground-truth sign which we will refer to as $B^{(i)}$. The truth $T^{(i)}$ is the network prediction for sample $i$. Lastly, the false value $S^{(i)}$ is the ground-truth for any other sample ${j\neq i \in \{B\}}$, where $j$ is randomly selected.

\subsection{Loss functions for training SignNet}
\subsubsection{L$_2$ Regression loss ($\mathcal{L}_a$) }
The objective here is to learn  the   probability $p(V|S)$  of  producing  a sequence of sign-poses $V= (s_1, \ldots, s_T)$ over $T$ time steps, given a spoken/written language sentence $S= (w_1, \ldots, w_U)$ having $U$ words. 

Given the text sentence $S$ as the inputs, the completed decoder output sequence of pose-signs can be expressed as $\hat{s}_{1:T} = \hat{s}_1,\ldots, \hat{s}_T$. The Mean Squared Error (MSE) loss between the predicted sequence, $\hat{s}_{1:T}$, and the ground truth,${s}_{1:T}^T$ is given as:
\begin{eqnarray}
\mathcal{L}_a = \mathcal{L}_{MSE} = \frac{1}{T}\sum_{i=1}^t (s^T_{1:T} - \hat{s}_{1:T})^2 
\end{eqnarray}

\begin{figure*}[ht]
	\centering
	\includegraphics[width=.85\textwidth]{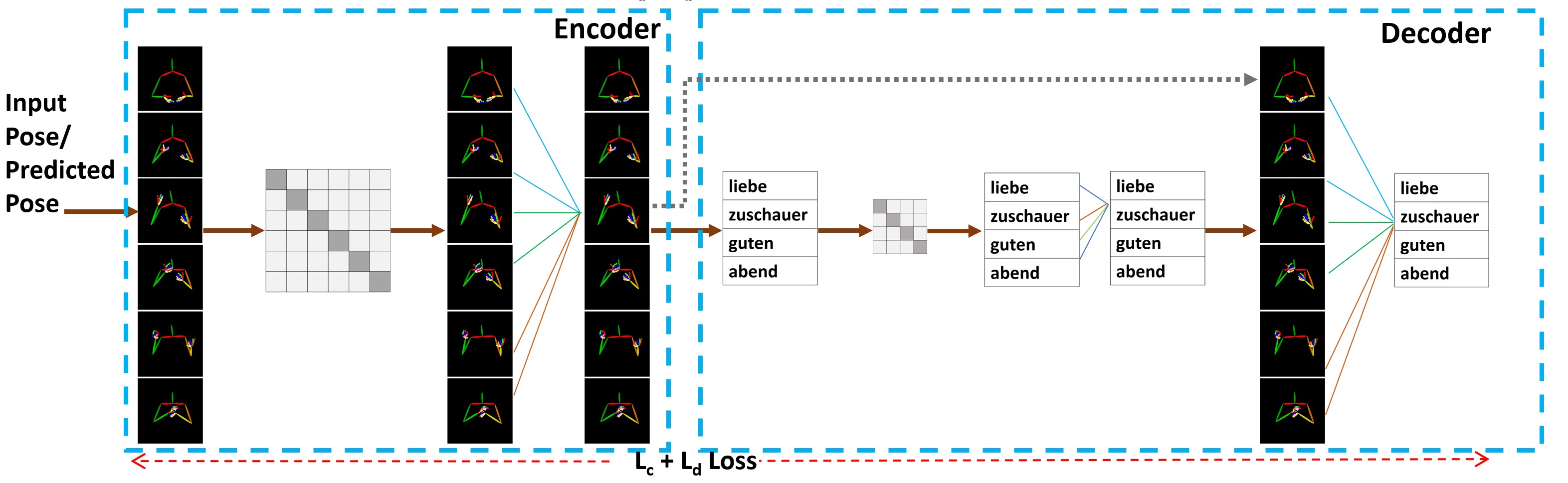} 
	\caption{SignNet pose-to-text (for evaluation): $L_{c}, and L_{d}$ are the losses used. Details in Section \ref{method:signn} (Best viewed in color).}
	\label{fig:signnet_eval}
\end{figure*}

\subsubsection{Sign similarity metric-based loss ($\mathcal{L}_b$)}
As shown in Figure \ref{fig:triplet}, $L_b$ is introduced to reduce the network confusing similar signs. 
The sign similarity based loss over all $M$ samples can thus be given as:
\begin{eqnarray}
\mathcal{L}_b = \sum_i^M d(B^{(i)},T^{(i)},S^{(i)})
\label{eqn:triplet4}
\end{eqnarray}
\noindent \underline{Justification:} There is only a finite number of valid poses that make up a sign, hence there is often significant overlap between signs in the same batch. Without the strong constraint to separate truth and false examples, the network confuses signs when generating from text.

\subsubsection{Total loss: }
SignNet for text-to-pose generation is trained to improve the pose generation by using a weighted combination of the regression and metric-based losses:
\begin{eqnarray}
\label{eq:total_ph1}
\mathcal{L}_{Text2Pose} = \lambda_a L_a + \lambda_b L_b. 
\end{eqnarray}

In our experiments, using a grid search, we obtained the best performance with $\lambda_a=5$ and $\lambda_b=5$.  Additional ablations performed on the SignNet model are shown in Table \ref{tab:t2pabl}

\section{The Evaluation Model (Pose-to-Text)}
The poses generated from the first phase of SignNet are evaluated in a sign-to-text translation network as shown in Figure \ref{fig:signnet_eval}. In this network, the encoder is trained to learn the inter-dependencies between the different pose sequences from the input sign language video, where the inpus to the encoder is a sequence of poses. The decoder learns the dependencies between different words, and also between words and pose sequences to provide efficient text translation. The poses generated from SignNet (text to pose ) are dumped to file during test time and they are then passed through this evaluation network to obtain their text translations.

\subsection{Loss functions for training the evaluation model}

\subsubsection{Recognition loss ($\mathcal{L}_c$)}
\paragraph*{Gloss definition:} \emph{Gloss} is the written set of notations used to transcribe sign language into its written/spoken counterpart. Given that sign language is a visual-spatial language, in the absence of videos, gloss notations can be used to capture  the sign-for-sign word ordering. It does this by providing different symbols useful for representing the spatial-temporal, facial and 3D body grammar present in sign language. The grammar of sign language is very different from that of its spoken/written language counterpart, but the use of gloss alleviates this problem.

\paragraph{Gloss-driven recognition:} Hence, we train the evaluation model first using the recognition loss.  Given the input sign video, as a sequence of sign poses $V=(s_1, \ldots, s_T)$ and the sequence of glosses $G= (g_1, \ldots, g_N)$ corresponding to $V$, the goal here is to learn $p(G|V)$. Because this sign to gloss mapping is monotonic and the word orderings are relatively consistent between signs and glosses, though requiring alignment between sequences of varied lengths, we employ the connectionist temporal classification loss, or CTC loss \cite{graves2006ctc} for SL recognition. 

CTC computes the loss between the unsegmented stream of input sign-pose embeddings and the target sequence of glosses. First, we obtain the pose-level gloss probabilities $p(g_t|V)$ by projecting the embeddings through a linear layer and softmax activation.

Then if we consider a path $\pi= (\pi_1, \cdots, \pi_T)$,  the probability of a viable path given the video can be written as $p(\pi|V)$. CTC is thus used marginalize over the possible alignments of $V$ to $G$, such that:

\begin{eqnarray}
p_{ctc}(G|V) = \sum_{\pi \in \mathcal{M}} p(\pi|V) 
\end{eqnarray}

\noindent where $\mathcal{M}$ is the set of viable paths in the sequence $G$; \\
\noindent Hence,
\begin{eqnarray}
\mathcal{L}_c = \mathcal{L}_{ctc}  = 1 - p(G^T|V)
\end{eqnarray}
\noindent where $G^T$ is the ground-truth gloss sequence corresponding to video $V$.

\subsubsection{Translation loss ($\mathcal{L}_d$)}
The primary task of the evaluation model is to generate a written/spoken language sentence $S = (w_1,\ldots,w_U)$ given a sign video $V$, as defined previously. The translation process discussed here aims to learn $p(S|V)$. Going from pose to sentences, in the decoding phase, we have:
\begin{eqnarray}
p(S|V) = \prod_{i=1}^U p(w_i | w_{i-1}) = \prod_{i=1}^U \mathbf{Z}_{i, s_i}
\end{eqnarray}
\noindent where $U$ is the length of the sentence and $\mathbf{Z} = (Z_{j,k}) = [z_1, \ldots, z_U]^\top$ is the probability distribution of the sentence when translated; $Z_{j,k}$ is the probability of word $w_j$ having a word label $k$, given $w_{j-1}$. The loss function is therefore:
\begin{eqnarray}
\mathcal{L}_d   = 1 - p(S^T|V)
\end{eqnarray}
\noindent where $S^T$ is the ground truth sentence corresponding to video $V$, comprising of the aggregation of the ground truth probability of words during the decoding phase.

\subsubsection{Pose-to-text total loss($\mathcal{L}_{Pose2Text}$)}
The evaluation model maximizes its performance by using a weighted combination of the recognition and translation losses for text translation:
\begin{eqnarray}
\label{eq:total_ph2}
\mathcal{L}_{Pose2Text} = \lambda_c L_c + \lambda_d L_d. 
\end{eqnarray}

\noindent The best performance of the evaluation model was obtained with $\lambda_c=100$ and $\lambda_d =100$ (multiplying the losses by a factor of 100 improves the performance, probably by boosting the gradients to be propagated).

\section{Experiments and Results}
\begin{figure*}[htp]
	\centering
	\includegraphics[width=0.90\textwidth]{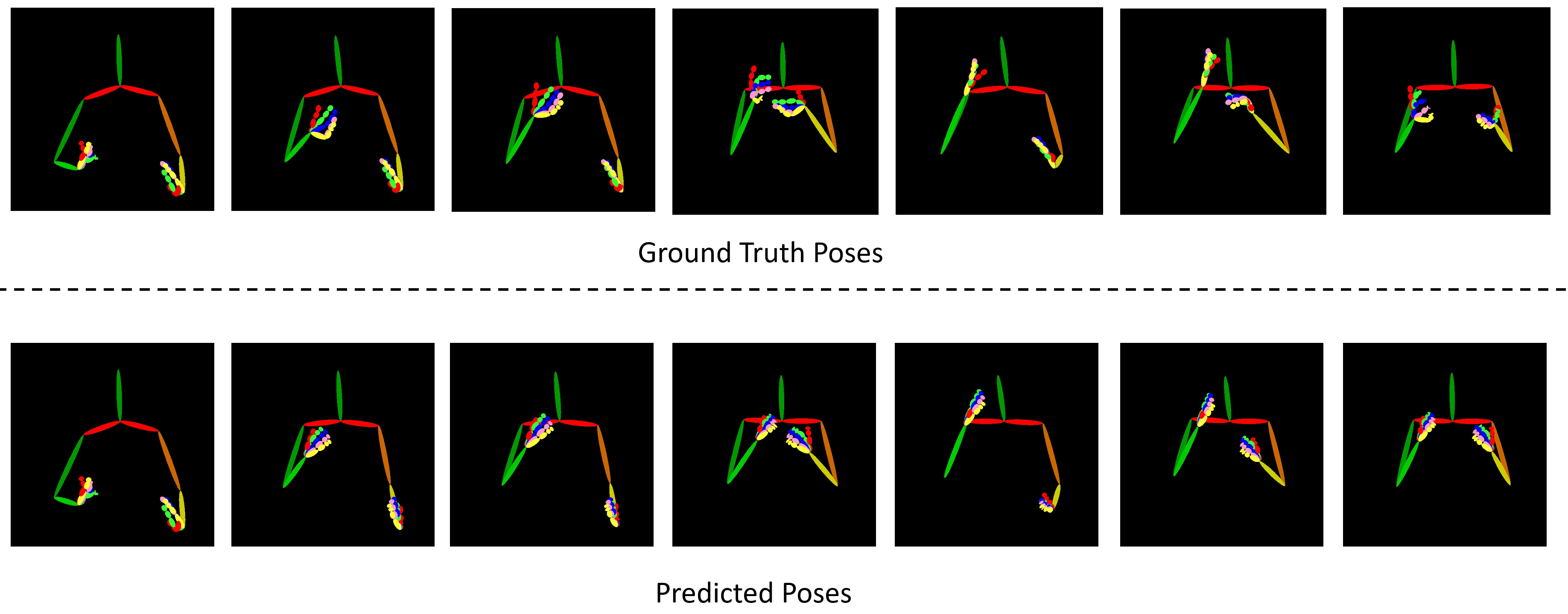} 
	\caption{Ground Truth (Top) and Predicted (Bottom) poses using metric embedded learning. Test samples with lower loss and same time frames are chosen for display.}
	\label{fig:predictedpose}
\end{figure*}

\begin{table*}[t]
	\centering
	\begin{tabular}{|p{6.0cm}|p{6.0cm}|}
		\hline
		Ground Truth & Predicted \\ \hline
		G: liebe zuschauer guten abend & G: liebe guten abend \\ \cdashline{1-2}
		E: dear viewers good evening & E: love good evening \\ \hline
		G: im süden freundliches wetter & G: im weht wetter \\ \cdashline{1-2}
		E: in the south friendly weather & E: in the blowing weather \\ \hline
		G: und nun die wettervorhersage für morgen sonntag den fünften dezember & G: und die wettervorhersage für morgen donnerstag den achtzehnten juli \\ \cdashline{1-2}
		E: and now the weather forecast for tomorrow sunday the fifth of december & E: and the weather forecast for tomorrow thursday the eighteenth of july \\ \hline
		G: der wind weht meist schwach aus unterschiedlichen richtungen & G: im weht schwach schwach aus unterschiedlichen richtungen \\ \cdashline{1-2}
		E:the wind usually blows weakly from different directions & E: im blowing weak weak from different directions \\ \hline
		G: sonst ein wechsel aus sonne und wolken & G: im seltener aus sonne und wolken \\ \cdashline{1-2}
		E:otherwise an alternation of sun and clouds & E: in the seldom sun and clouds \\ \hline
	\end{tabular}
	\caption{Examples of our outputs: showing the German ground truth texts (left column) and the predicted texts (right column), obtained by passing the predicted poses from SignNet through the sign to text translation evaluation network. The subsequent English translations for both the ground truth texts and predicted ones are also displayed (G: German, E: English). The Google German-to-English translator was used.}
	\label{tab:german_english}
\end{table*}

\subsection{Dataset and Metrics}
We evaluate SignNet on 
the RWTH-PHOENIX-Weather benchmark dataset (RWTH) and use the same split (7096/519/642, train/dev/test) as provided by the original authors\cite{RWTH_dataset}. 

To evaluate our SignNet pose-to-text translation we use BiLingual Evaluation Understudy (BLEU) \cite{bleu} metric that evaluates how good the translation is by comparing the predicted text to its ground truth equivalent. BLEU 1 - BLEU 4 scores evaluate the performance based on 1-gram (individual words) to 4-gram words (group of consecutive four words). Additionally, we evaluate our text-to-pose network using the dynamic time warping (DTW) \cite{dtw} algorithm. We use DTW to help align the poses to as close to ground truth as possible. 

While training SignNet for efficient pose generation we use DTW as our evaluation metric and optimize the network for the lowest DTW score. For the evaluation network BLEU is used as the evaluation metric for the pose-to-text translation.

\subsection{Results}

\subsubsection{Predictions and ground-truth texts}
In Table \ref{tab:german_english}  we list some of our best translations of the generated pose (based on high BLEU scores).

\subsubsection{Predictions and ground-truth poses}
We select one of the best generated pose sequences (based on the lowest DTW costs) obtained from SignNet, and show in Figure \ref{fig:predictedpose} how our predicted pose closely aligns with the ground truth. 

\subsubsection{Comparisons with other methods}
In Table \ref{tab:pose_out}, we compare the performance of our SignNet translations for generated poses with the single channel poses generated by recent sign language production mechanisms, as well as a baseline method using 2D OpenPose features. Although several multichannel mechanism have been presented in the literature, in this work, we focus on single channel.
\begin{table*}[]
	\centering
	\begin{tabular}{|l|c|c|c|c|}
		\hline
		Experiment type & \multicolumn{4}{c|}{Test} \\ \hline
		& BLEU 1 & BLEU 2 & BLEU 3 & BLEU 4 \\ \hline
		PT G2P \cite{progressivetransformer}                  
		& 31.8 & 19.19 & 13.51 & 10.43  \\ \hline
		PT T2P \cite{progressivetransformer}                  
		& 31.36 & 19.04 & 13.54 & 10.51 \\ \hline
		2D Pose (using SignNet)                    
		& 23.42 &13.13 & 8.75 & 6.32  \\ \hline
		With facial expressions (2022)\cite{viegas2022including}                   
		& 27.76 &18.86 & 14.11 & 11.32  \\ \hline 
		Mixture of Motion Primitives\cite{saunders2021mixed}                   
		& 35.89 &23.27 & \textbf{16.86} & \textbf{13.30}  \\ \hline 
		
		SignNet G2P (ours)                                
		& \textbf{39.14}   & \textbf{23.98}   & 16.41   & 11.84 \\ \hline
		SignNet T2P* (ours)                                  
		& \textbf{36.76}   & \textbf{21.78}   & \textbf{14.77}   & \textbf{10.66}\\ \hline
	\end{tabular}
	\caption{Translation performance on predicted poses using SignNet. G2P - Gloss-to-Pose, T2P - Text-to-Pose. *Note in the last row that the network takes in texts and generates sign poses.}
	\label{tab:pose_out}
\end{table*}

\begin{table*}[]
	\centering
	\begin{tabular}{|p{6.0cm}|c|c|c|c|}
		\hline
		Experiment type & \multicolumn{4}{c|}{Test} \\ \cline{2-5}
		& BLEU1 & BLEU2 & BLEU3 & BLEU4 \\ \hline
		SignNet G2P (ours w/o metric-based loss) & 36.41 & 20.97 & 13.52 & 9.04		 \\ \hline
		SignNet T2P (ours w/o metric-based loss) &	36.19 &  20.77 & 13.27 & 8.8 \\ \hline	
		
		SignNet G2P (ours w/ metric-based loss)        & \textbf{39.14}   & \textbf{23.98}   & \textbf{16.41}   & \textbf{11.84} \\ \hline
		SignNet T2P (ours w/ metric-based loss)      & \textbf{36.76}   & \textbf{21.78}   & \textbf{14.77}   & \textbf{10.66}\\ \hline
	\end{tabular}
	\caption{Text to sign translation results using SignNet with (w/) and without (w/o) metric-based loss for sign similarity.}
	\label{tab:metricloss}
\end{table*}

To verify that our metric embedded learning helps in improving the pose generation, we evaluate our model with and without metric embedded loss. Table \ref{tab:metricloss} shows that SignNet with metric embedding loss provides a good boost in performance by achieving higher BLEU 1 - BLEU 4 scores.

Table \ref{tab:t2pabl} shows the results obtained from performing ablations, to determine the best performing loss combination weights when testing the SignNet model.

\begin{table}[]
	\centering
	\begin{tabular}{|l|c|c|c|c|c|c|}
		\hline
		Model & Rec. & Trans. & BLEU 1 & BLEU 2 & BLEU 3 & BLEU 4 \\ 
		& wt.  & wt. &  & &  &  \\ 
		& $(\lambda_a)$ &  ($\lambda_b$) &  & &  &  \\  \hline
		G2P               
		& 1 & 10 & 33.87 & 19.08 & 12.18 & 8.21 \\ \cdashline{1-7}[.4pt/1pt]
		T2P         
		& & & 33.61 & 18.74 & 12.01 & 8.07 \\ \hline
		G2P                   
		& 5 & 1 & 35.19 & 20.58 & 13.39 & 9.04 \\ \cdashline{1-7}[.4pt/1pt]
		T2P                   
		& & & 34.91 & 20.27 & 13.06 & 8.85  \\ \hline
		G2P                   
		& 5 & 5 & 39.14 & 23.98 & 16.41 & 11.84 \\ \cdashline{1-7}[.4pt/1pt]
		T2P                   
		& & & 36.76 & 21.78 & 14.77 & 10.66  \\ \hline
		G2P                   
		& 5 & 10 & 32.46 & 17.84 & 10.89 & 7.21 \\ \cdashline{1-7}[.4pt/1pt]
		T2P                   
		& & & 32.71 & 17.85 & 11.12 & 7.33 \\ \hline
		G2P                   
		& 10 & 5 & 33.73 & 18.66 & 11.44 & 7.44 \\ \cdashline{1-7}[.4pt/1pt]
		T2P                   	
		& & & 33.35 & 18.52 & 11.65 & 7.91 \\ \hline
	\end{tabular}
	\caption{Text to sign translation results using different values of recognition and translation loss weights}\label{tab:t2pabl}
\end{table}

\subsection{Optimization and implementation}
The input data used in training SignNet are 3D poses generated by extracting 2D openpose \cite{openpose_1} joints, then converting them to 3D poses \cite{Zelinka_2020_WACV}. 
We only chose joints from the head to the trunk and the ordering of the joints closely followed the standard Openpose \cite{openpose_1} structure. 

SignNet provided the best performance when using $2$ encoder and decoder layers. Our evaluation network performed the best with $7$ encoder layers and $2$ decoder layers (see Fig \ref{fig:signnet_eval})with an embedding dimension of 128.  We optimized SignNet using Adam optimizer \cite{adam} with a learning rate of $0.001$ and plateau scheduling.

\subsection{Discussion}
We perform various experiments to evaluate the performance of our SignNet network. Table \ref{tab:pose_out} highlights its pose generation capabilities. 
We first generate output poses on the dev set and test after training SignNet as shown in Figure \ref{fig:signnet}. After the poses are generated, we pass them back as input through the evaluation network (Figure \ref{fig:signnet_eval}). This allows us to determine how good our predicted poses are. 

We observe that our metric embedded learning boosts the performance of SignNet, achieving BLEU 1 score improvement from 31.80/31.36 (G2P/ T2P) $\rightarrow$ 39.14/36.76 (G2P/ T2P) through BLEU 4 score improvements from 10.43/10.51 $\rightarrow$ 11.84/10.66 (G2P/ T2P). To investigate how well the 2D pose points would perform compared with 3D points, we trained SignNet with 2D pose generated from OpenPose \cite{openpose_1} and compared it with the model trained on 3D pose. We observe that SignNet T2P trained on 3D points provides large improvements (BLEU 1: 23.42 $\rightarrow$ 36.76, BLEU 4: 6.32 $\rightarrow$ 10.66) when compared to the model trained on 2D poses.\medskip

\paragraph*{Limitations:} 
The main limitation of our proposed SignNet model is its exposure to a relatively confined  dataset whose subject domain is the weather and having a collective group of homogeneous signers.
Also, although we have developed a context-aware architecture, many aspects of sign language traits, especially those involving long-range temporal dependencies among signs are still not well understood in current-day automated SLT models. For example, it is not clear that a model such as proposed here can readily incorporate the unique but commonly-used linguistic spatial constructs such as the repeated use of spatial reference points, as described in the Section \ref{sec:intro}.\\

%
%

\section{Conclusion}
In this paper, we have presented SignNet, a text-to-sign interpretation model which combines several different losses, in particular a novel similarity metric-based loss which significantly improves our sign generation performance. 
Because sign language has many linguistic aspects that involve the 3D space around a signer, which create the context for ensuing signs, SignNet is a context-aware network, that has successfully learned temporal dependencies by co-relating all the frames in an input sequence to each other. 

We demonstrate that SignNet predicts  ``good''  signs when presented with input texts in scope. We demonstrate this by qualitatively examining predicted signs and we compare them with their ground-truth counterparts. Similarly, we qualitatively compare predicted texts with the corresponding ground truth. 

Although there is still much to explore in our future work as stated earlier, especially regarding the dual learning of sign-to-text and text-to-sign simultaneously, in this work, we have successfully presented a single channel text-to-sign language generation model,  partly useful for facilitating the two-way natural language communication between hearing and DHH individuals.


%


{\small
\bibliographystyle{ieee}
\bibliography{FG2023_txt2sgn}

\begin{thebibliography}{10}\itemsep=-1pt

\bibitem{bahdanau2016neural}
D.~Bahdanau, K.~Cho, and Y.~Bengio.
\newblock Neural machine translation by jointly learning to align and
  translate.
\newblock {\em arXiv preprint arXiv:1409.0473}, 2014.

\bibitem{dtw}
D.~J. Berndt and J.~Clifford.
\newblock Using dynamic time warping to find patterns in time series.
\newblock In {\em Proceedings of the 3rd International Conference on Knowledge
  Discovery and Data Mining (KDD)}, volume~10, pages 359--370. Seattle, WA,
  USA:, 1994.

\bibitem{RWTH_dataset}
N.~C. Camg{\"{o}}z, S.~Hadfield, O.~Koller, H.~Ney, and R.~Bowden.
\newblock Rwth-phoenix-weather 2014 t: Parallel corpus of sign language video,
  gloss and translation.
\newblock In {\em IEEE Conf. on Computer Vision and Pattern Recognition, Salt
  Lake City, UT, 2018}, CVPR, 05 2018.

\bibitem{openpose_1}
Z.~Cao, G.~Hidalgo, T.~Simon, S.~Wei, and Y.~Sheikh.
\newblock Openpose: Realtime multi-person 2d pose estimation using part
  affinity fields.
\newblock {\em {IEEE} Trans. Pattern Anal. Mach. Intell.}, 43(1):172--186,
  2021.

\bibitem{ebling-huenerfauth-2015-bridging}
S.~Ebling and M.~Huenerfauth.
\newblock Bridging the gap between sign language machine translation and sign
  language animation using sequence classification.
\newblock In {\em Proceedings of {SLPAT} 2015: 6th Workshop on Speech and
  Language Processing for Assistive Technologies}, pages 2--9, Dresden,
  Germany, Sept. 2015. Association for Computational Linguistics.

\bibitem{Glauert2006VANESSAA}
J.~Glauert, R.~Elliott, S.~Cox, J.~Tryggvason, and M.~Sheard.
\newblock Vanessa - a system for communication between deaf and hearing people.
\newblock {\em Technology and Disability}, 18:207--216, 2006.

\bibitem{graves2006ctc}
A.~Graves, S.~Fern{\'a}ndez, F.~Gomez, and J.~Schmidhuber.
\newblock Connectionist temporal classification: labelling unsegmented sequence
  data with recurrent neural networks.
\newblock In {\em Proceedings of the 23rd international conference on Machine
  learning}, pages 369--376, 2006.

\bibitem{huenerfauth2012effect}
M.~Huenerfauth and P.~Lu.
\newblock Effect of spatial reference and verb inflection on the usability of
  sign language animations.
\newblock {\em Universal Access in the Information Society}, 11(2):169--184,
  2012.

\bibitem{KARPOUZIS200754}
K.~Karpouzis, G.~Caridakis, S.-E. Fotinea, and E.~Efthimiou.
\newblock Educational resources and implementation of a greek sign language
  synthesis architecture.
\newblock {\em Computers \& Education}, 49(1):54--74, 2007.
\newblock Web3D Technologies in Learning, Education and Training.

\bibitem{adam}
D.~P. Kingma and J.~Ba.
\newblock Adam: A method for stochastic optimization.
\newblock {\em arXiv preprint arXiv:1412.6980}, 2014.

\bibitem{10.1007/978-3-642-23974-8_13}
M.~Kipp, A.~Heloir, and Q.~Nguyen.
\newblock Sign language avatars: Animation and comprehensibility.
\newblock In H.~H. Vilhj{\'a}lmsson, S.~Kopp, S.~Marsella, and K.~R.
  Th{\'o}risson, editors, {\em Intelligent Virtual Agents}, pages 113--126,
  Berlin, Heidelberg, 2011. Springer Berlin Heidelberg.

\bibitem{KOLLER2015108}
O.~Koller, J.~Forster, and H.~Ney.
\newblock Continuous sign language recognition: Towards large vocabulary
  statistical recognition systems handling multiple signers.
\newblock {\em Computer Vision and Image Understanding}, 141:108--125, 2015.
\newblock Pose \& Gesture.

\bibitem{McDonald2015AnAT}
J.~C. McDonald, R.~J. Wolfe, J.~Schnepp, J.~Hochgesang, D.~Jamrozik, M.~Stumbo,
  L.~Berke, M.~Bialek, and F.~Thomas.
\newblock An automated technique for real-time production of lifelike
  animations of american sign language.
\newblock {\em Universal Access in the Information Society}, 15:551--566, 2015.

\bibitem{bleu}
K.~Papineni, S.~Roukos, T.~Ward, and W.-J. Zhu.
\newblock Bleu: A method for automatic evaluation of machine translation.
\newblock In {\em Proceedings of the 40th Annual Meeting on Association for
  Computational Linguistics}, ACL '02, pages 311--318, Stroudsburg, PA, USA,
  2002. Association for Computational Linguistics.

\bibitem{DBLP:journals/corr/abs-2103-15910}
R.~Rastgoo, K.~Kiani, S.~Escalera, and M.~Sabokrou.
\newblock Sign language production: {A} review.
\newblock {\em CoRR}, abs/2103.15910, 2021.

\bibitem{SaundersBMVC20}
B.~Saunders, R.~Bowden, and N.~C. Camg{\"{o}}z.
\newblock Adversarial training for multi-channel sign language production.
\newblock In {\em 31st British Machine Vision Conference 2020, {BMVC} 2020,
  Virtual Event, UK, September 7-10, 2020}. {BMVA} Press, 2020.

\bibitem{progressivetransformer}
B.~Saunders, N.~C. Camg{\"{o}}z, and R.~Bowden.
\newblock Progressive transformers for end-to-end sign language production.
\newblock {\em CoRR}, abs/2004.14874, 2020.

\bibitem{saunders2021mixed}
B.~Saunders, N.~C. Camgoz, and R.~Bowden.
\newblock Mixed signals: Sign language production via a mixture of motion
  primitives.
\newblock In {\em Proceedings of the IEEE/CVF International Conference on
  Computer Vision}, pages 1919--1929, 2021.

\bibitem{progressive}
B.~Saunders, N.~Cihan, Camg{\"{o}}z, and R.~Bowden.
\newblock Continuous 3d multi-channel sign language production via progressive
  transformers and mixture density networks.
\newblock {\em CoRR}, abs/2103.06982, 2021.

\bibitem{FaceNet}
F.~Schroff, D.~Kalenichenko, and J.~Philbin.
\newblock Facenet: A unified embedding for face recognition and clustering.
\newblock In {\em Proceedings of the IEEE conference on computer vision and
  pattern recognition}, pages 815--823, 2015.

\bibitem{transformer}
A.~Vaswani, N.~Shazeer, N.~Parmar, J.~Uszkoreit, L.~Jones, A.~N. Gomez,
  L.~Kaiser, and I.~Polosukhin.
\newblock Attention is all you need.
\newblock {\em CoRR}, abs/1706.03762, 2017.

\bibitem{viegas2022including}
C.~Viegas, M.~Inan, L.~Quandt, and M.~Alikhani.
\newblock Including facial expressions in contextual embeddings for sign
  language generation.
\newblock {\em arXiv preprint arXiv:2202.05383}, 2022.

\bibitem{Zelinka_2020_WACV}
J.~Zelinka and J.~Kanis.
\newblock Neural sign language synthesis: Words are our glosses.
\newblock In {\em Proceedings of the IEEE/CVF Winter Conference on Applications
  of Computer Vision (WACV)}, March 2020.

\end{thebibliography}
}

\end{document}